\documentclass[a4paper,11pt]{article}
\pdfoutput=1 

\usepackage{jcappub} 

\usepackage[T1]{fontenc} 
\usepackage{lscape}
\usepackage{graphicx}
\usepackage{epsfig}
\usepackage{ulem}

\title{\boldmath Qualitative analysis for viscous cosmologies in a non linear regime of the Israel-Stewart formalism}

\author[a,b]{Gilberto Aguilar-P\'erez\footnote{Corresponding author.}}
\author[a,b]{Ana A. Avilez-López}
\author[c]{Miguel Cruz}
\affiliation[a]{Facultad de Ciencias F\'{\i}sico Matem\'aticas, Benem\'erita Universidad Aut\'onoma de Puebla, Apdo. Postal 1152, Puebla, Pue., M\'exico.}
\affiliation[b]{Centro Internacional de F\'isica Fundamental, Benem\'erita Universidad Aut\'onoma de Puebla, Apdo. Postal 1152, Puebla, Pue., M\'exico. }
\affiliation[c]{Facultad de F\'{\i}sica, Universidad Veracruzana 91000, Xalapa, Veracruz, M\'exico}

\emailAdd{gilberto.aguilarp@alumno.buap.mx}
\emailAdd{ana.avilezlopez@correo.buap.mx}
\emailAdd{miguelcruz02@uv.mx}

\abstract{We explore the dynamical properties of a cosmological model that includes viscous effects in the dark matter sector of the fluid equations in a flat Friedmann-Lemaitre-Robertson-Walker (FLRW) spacetime. The bulk viscous effects are described by a non linear extension of the full Israel-Stewart model, which is a fluid causal scheme. We allow the interchange of energy in the dark sector and describe this by means of the interaction term, namely $Q$. We establish the dynamical system corresponding to Friedmann and fluid set of equations associated to the model and study the linear stability of its critical points. From the exploration of the dynamical system, we show the appearance of a critical point characterizing a de Sitter universe within the non interacting and interacting dark sector. {We focus our study to analyse the stability of this fixed point in a large region of parameter space and derive linearized solutions around it. These approximate and analytical solutions are potentially able to describe the expansion of the universe since they are close to a de Sitter stationary solution}. Within this regime with $Q \neq 0$, we realize the existence of regions in the space of parameters where this critical point is stable and describes the behavior of dark energy as quintessence, cosmological constant and phantom like fluids. We perform a comparison between numerical and linearized solutions nearby the critical points within the full non linear regimes and also contrast them against $\Lambda$CDM model as a fiducial model. We find that the fully non linear regime is favored by observations and closer to the concordance model due to the non-zero value of the parameter $j$, which controls the non linear effects of bulk viscosity. In fact, at low redshift values, the expansion rate associated to the full non linear regime is practically indistinguishable from the $\Lambda$CDM model. The deceleration parameter obtained in this regime exhibits a transition from decelerated to accelerated cosmic expansion.}

\begin{document}
\maketitle
\flushbottom

\section{Introduction}
\label{sec:intro}
Most of the processes involved in the evolution of our universe can be described in good approximation by assuming that matter content behaves as a perfect fluid. However, this assumption does not take into account some physical aspects such as heat transfer or changes in the entropy (to mention some), therefore the picture is incomplete. Despite the adequate results obtained with the perfect fluid description, astrophysics and cosmology require of a more detailed description for the fluid dynamics in order to understand our observable universe. For instance, recent works show that the growth of entropy could explain the current accelerated expansion of the universe \cite{growth1, growth2}\footnote{It is worthy to mention that in fluid cosmology the most simple proposal for non vanishing entropy growth is given by bulk viscosity \cite{entro}.}; to our knowledge, the identity of the catalyst for the acceleration of the cosmic evolution remains in mystery but the astrophysical observations indicate that we can not neglect its existence \cite{data1, data2, data3, data4}.\\

In the realm of cosmology a direct way to consider the already mentioned physical aspects in the fluid description is given by the inclusion of bulk viscosity, which is denoted usually as $\Pi$ or termed simply as bulk pressure. The origin of bulk viscosity, as shown in Ref. \cite{origin}, can be attributed to different temperature evolution of the subsystems that make up the universe, this is compatible with the single fluid description. If the universe is described by a fluid as a whole, then the total particle number density, $n$, can be decomposed in terms of each component. In fact, the origin of bulk viscosity is independent of the interaction that could exist between the different components. However, the introduction of bulk viscosity in the cosmological equations requires of the grounds of irreversible thermodynamics. A first attempt to construct a theory of irreversible thermodynamics consistent with General Relativity was made by Eckart \cite{eck}, but it was found that the velocity of bulk viscous perturbations is superluminal, i.e., non causal. The causal framework for bulk viscosity was proposed in \cite{is}, this scheme is usually known as the Israel-Stewart theory, but we must have in mind that such approach is valid only in the near equilibrium condition, 
\begin{equation}
    \left|\frac{\Pi}{p}\right| \ll 1,
    \label{eq:near}
\end{equation}
being $p$ the local equilibrium pressure of the fluid. Nowadays we can find a lot of works where the Israel-Stewart and Eckart models are implemented to describe the late times behavior of the universe and tested against observations, see for instance \cite{obs1, obs2, obs3, obs4, obs5, obs6, obs7}\footnote{The role of bulk viscosity have been also explored in the context of large scale structures formation, in Ref. \cite{large} can be found that the clustering properties of viscous dark matter are consistent with the structures observed in our universe.}. However, the condition (\ref{eq:near}) it is not satisfied in the case of an expanding fluid \cite{cqg95, maartens}. A more satisfactory model for bulk viscosity can be found in Ref. \cite{mendez}, where the authors propose a non linear extension of the Israel-Stewart model in order to relax the condition (\ref{eq:near}) and apply it to expanding fluids but at the same time they maintain a causal scheme; therefore we consider that this approach is more adequate to describe the behavior of the observable universe at late times than other scenarios. In Ref. \cite{causal} it was found that the causality of the Einstein equations coupled to a non linear version of the Israel-Stewart model is ensured, in fact, the authors provide the basis to include bulk viscosity effects in numerical simulations for gravitational waves.\\

The main goal in this work is to study the lineal stability of cosmological solutions corresponding to models in which viscous fluid components are described by the non linear extension of the Israel-Stewart formalism. For this purpose, we use the so called qualitative analysis of autonomous dynamical systems firstly introduced by Liapounov and Poincaré \cite{Coley, BoyceDiPrima}. As first step we extend the results of Ref. \cite{lepesys}, where the dynamical analysis of the Israel-Stewart model was performed in the linear regime; then we shall compare our results with the $\Lambda$CDM model. Another characteristic of our cosmological model is that we will allow interactions in the dark sector, this is, the interchange of energy between the viscous dark matter and dark energy is not forbidden. The interacting scheme have been considered for bulk viscosity previously, see for instance Ref. \cite{yoelsy}. 

By following the prescription of the qualitative approach in the dynamical analysis, the stability of the stationary solutions of the dynamical system is studied and also linearized solutions are obtained around a specific critical point with particular physical relevance. As we will see below, the behavior of these linearized solutions leads to an interesting conclusion:  the full non linear description seems to be a more adequate proposal to describe the expansion rate of our universe rather than the linear solutions even at high redshift. Although the non linear description of bulk viscous effects implies a great complexity, we can have a global perspective of this cosmological model through the dynamical system approach. Nevertheless, we also computed exact numerical solutions in order to test the validity of linearized solutions.

The global structure of bulk viscous cosmologies was also discussed earlier in Ref. \cite{qual1, qual2} (and references therein). Other works in this sense can be found in \cite{qual3, qual4}.\\

The plan of this work is the following: in Section \ref{sec:isform} we provide a brief description of the non linear regime of the Israel-Stewart model and we also present {the cosmological model, that is, the specific} set of differential equations that will govern the fluid {and the scale factor} description. In Section \ref{sec:dym_sys} we describe the dynamical system to be studied. We find the critical points of the dynamical system and explore their stability for the non interacting and interacting dark sector cases. We focus on a critical point that characterizes a de Sitter universe and we study the linear stability of this point in the space of parameters. In section \ref{sec:linearized} we find analytical  solutions of the linearized dynamical system around the critical point and compute some relevant cosmological {quantities} for the model, such as the Hubble rate and the deceleration parameter. We compare {these quantities derived from  analytical solutions} with {those coming from the full dynamical system} and those corresponding to the $\Lambda$CDM model. Section \ref{sec:conclusions} is devoted to the final comments of our work. In appendix \ref{app:stabi} we present details about the stability conditions for the critical point of interest. In this work we will consider $8\pi G=c=k_{B}=1$ units.

\section{The non linear Israel-Stewart formalism setup in FLRW universes}
\label{sec:isform}
In our description we will consider an homogeneous and isotropic viscous expanding fluid in a spatially flat FLRW spacetime, described by the following differential line element
\begin{equation}
    ds^2=-dt^2+a^2(t)\left(dx^2+dy^2+dz^2\right).
\end{equation}

Where $a(t)$ es the only free function usually dubbed as the scale factor.
The fluid is characterized by an energy density $\rho(t)$ and pressure $p_{\mathrm{eff}}(t)$. In general, the bulk viscosity effects on the fluid are expected to contribute on the equilibrium pressure, $p$, therefore we have
\begin{equation}
p_{\mathrm{eff}}=p+\Pi,    
\label{eq:effective_pressure}
\end{equation}
where $\Pi$ denotes the bulk viscous pressure. We will restrict ourselves to a barotropic EoS for the equilibrium pressure of each species, i.e., $p_{i}=\omega_{i} \rho_{i}$, being $\omega_{i}$ a constant parameter usually termed as parameter state which lies in the interval $0 \leq \omega < 1$ as long as the strong energy principle is satisfied. This parameter is usually written alternatively as, $\omega = \gamma - 1$. In this case, the Friedmann equations are given as follows  
\begin{align}
    & 3H^2 = \rho,\label{eq:Friedmann1} \\ 
   & \dot{H}+H^2 = -\frac{1}{6}(\rho+3p_{\mathrm{eff}}), \label{eq:Friedmann2}
\end{align}
where $H:= \dot{a}/a$ denotes the Hubble rate and quantifies the expansion of the universe as function of time. In our notation we will use the dot to denote derivatives w.r.t. cosmic time. From the acceleration equation (\ref{eq:Friedmann2}) and a barotropic EoS we can perform straightforward calculation to obtain the following expression, $\Pi = -2\dot{H} - 3H^{2}(1+\omega)$. Notice that the negativity of $\Pi$ is the reason why bulk viscous effects contribute to give rise to the accelerated expansion of our universe. Using the Eqs. (\ref{eq:Friedmann1}) and (\ref{eq:Friedmann2}) together with a barotropic EoS, the continuity equation for the energy density takes the form
\begin{equation}
    \dot{\rho} + 3H[(1+\omega)\rho + \Pi] = 0.
\end{equation}
Besides, in the non linear extension of the Israel-Stewart model, the bulk viscous pressure obeys the following evolution equation \cite{mendez}
\begin{equation}
\tau \dot{\Pi}\left(  1+\frac{\tau_{*}}{\zeta}\Pi  \right) + \Pi(1+3\tau_{*}H) = -3\zeta H -\frac{\epsilon}{2}\tau \Pi \left[ 3H+\frac{\dot{\tau}}{\tau}-\frac{\dot{\zeta}}{\zeta}-\frac{\dot{T}}{T}  \right]\left(  1+\frac{\tau_{*}}{\zeta}\Pi  \right), \label{eq:transporte}
\end{equation}
where $\tau$ is the relaxation time associated to viscous effects and is defined as, $\tau := \frac{\zeta}{v^2(\rho+p)}$, being $v$ the dissipative contribution to the speed of sound, $V$, coming from the bulk viscous perturbations \cite{maartens} and $\zeta$ corresponds to the bulk viscosity coefficient. With the purpose of preserving causality, $V$ must obey the following condition: $V^{2} = v^{2} + c^{2}_{s} \leq 1$, where the adiabatic contribution is defined as $c^{2}_{s} := (\partial p/\partial \rho)_{S}$. The characteristic time for non linear effects is given by $\tau_{*}=j^2\tau$, where $j$ is a constant parameter. It is worthy to mention that the limit case $j=0$ in Eq. (\ref{eq:transporte}) leads to the usual transport equation of the linear regime within the Israel-Stewart formalism \cite{is, is2, is2a, is2b, is3, cqg95}. Within this limit the value $\epsilon=0$ is known as truncated Israel-Stewart formalism and $\epsilon = 1$ is the full theory. In this work we will consider this latter value for $\epsilon$. Besides, for $\tau\rightarrow 0$ we observe that velocity of propagation of bulk viscous perturbations becomes large, therefore the model turns unstable and does not obey the aforementioned causality condition, this is known as Eckart model; we will discard this case from our description. For the bulk viscosity coefficient we will assume the usual power-law form
\begin{equation}
    \zeta = \zeta_{0}\rho^{s},
    \label{eq:zeta}
\end{equation}
where $\zeta_{0}$ and $s$ are positive constants. The positivity of $\zeta$ will guarantee an increasing behavior for the expansion rate of the universe within the viscous cosmology approach and a well defined cosmological model from the thermodynamics point of view, see Ref. \cite{zeta}. In contrast to the standard fluid approach used in most of the cosmological models, in a viscous fluid we have entropy production and heat dissipation, consequently, the temperature does not scale as the inverse of the scale factor, $T(a)=T_0/a$, rather it is determined by using the Gibbs integrability condition as function of the energy density as follows \cite{maartens} 
\begin{equation}\label{eq:T_rho}
T(\rho)= T_0\rho^{\omega /(1+\omega)}.
\end{equation}
Then Eq. (\ref{eq:transporte}) will be the fundamental dynamical equation for non linear bulk viscosity in a flat FLRW universe, which depends on the definitions provided for $T$, $\tau$ and $\zeta$.

\subsection{The Cosmological Models}
In this section we briefly describe the cosmological models to be analyzed. We will consider an universe in which the energy density budget is dominated by dark matter and dark energy, considering that baryons make up about 5 percent of the total amount of non-relativistic matter in the late universe, we neglect their contribution. We also ignore contributions of other components like massive neutrinos and radiation since they are not dynamically relevant at late times. In view of the unknown nature of dark matter and dark energy in this work we consider a viscous dark matter sector described by a non linear Israel-Stewart formalism, as can be seen in Eq. (\ref{eq:Cont2}). Besides, given the dominance of both components in the universe, it is equally reasonable that these dark components could interact among themselves, therefore a function denoted by $Q$ is needed to characterize the interaction in the dark sector. Together with the transport equation (\ref{eq:transporte}) for the viscous pressure, the system of differential equations governing the dynamics of the universe in a flat FLRW cosmology will be given by the Friedmann equation and the continuity equations for each specie, yielding
\begin{align}
	& 3H^2 = \rho_m+\rho_{DE},\label{eq:Fried2} \\
    & \dot{\rho}_m = -3H\rho_m-3H\Pi +Q,\label{eq:Cont2}\\
    & \dot{\rho}_{DE} = -3H(1+\omega_{DE})\rho_{DE} -Q,\label{eq:Cont3}
\end{align}
where, the total energy density is given by $\rho_T=\rho_m+\rho_{DE}$ with $\rho_m$ and $\rho_{DE}$ being the energy densities of dark matter and dark energy, respectively. Note that we focus on the cold dark matter case, i.e., $\omega_{m}=0$; its contribution to the total pressure of the fluid will be given only by $\Pi$, that obeys equation (\ref{eq:transporte}). In order to reduce the number of free parameters, for the bulk viscosity coefficient (\ref{eq:zeta}) we will take the value, $s=1/2$ and the functional form, $\zeta = \zeta(\rho_{m})$. Usually the value $s=1/2$ is assumed for simplicity, the form of (\ref{eq:transporte}) can be simplified in order to integrate it but this value also leads to several physical cases of interest in viscous cosmology, for instance the emergence of a phantom evolution in the linear and non linear approximations of Israel-Stewart theory \cite{lepe1, lepe2} and under certain conditions such value concedes a viscous cosmological model with accelerated late expansion that begins from an initial singularity \cite{is3}, see also \cite{cataldo}; specifically within the non linear regime for such value the use of observations revealed a viable scenario to explain the recent accelerated expansion of the universe without the addition of any exotic component, see Ref. \cite{nonlinear}. For the total energy density can be found the following continuity equation
\begin{equation}
 \dot{\rho_T}+3H\gamma_{\mathrm{eff}}\rho_T=0,   
\end{equation}
which resembles the standard equation obtained from the conservation condition of the energy-momentum tensor, $\nabla_{\mu}T^{\mu \nu} = 0$, note that this can be achieved despite the presence of the $Q$-term in the dark sector. The effective equation of state obeys, $\omega_{\mathrm{eff}}=p_{\mathrm{eff}}/\rho_T=(p_{DE}+\Pi)/\rho_T$.

\section{Dynamical system}
\label{sec:dym_sys}
As discussed in reference \cite{mendez}, in order to describe the cosmic evolution of an universe in which dissipative effects are described in a non linear extension of the Israel-Stewart formalism, we must deal with the transport equation given in Eq. (\ref{eq:transporte}) for the bulk viscous pressure and from there it is possible to obtain a non linear second order differential equation for the Hubble parameter, which is hard to solve. If we follow the standard procedure, two initial conditions are needed to obtain a well established solution, an initial value corresponding to the Hubble constant (if we take present time as initial value) and a second initial value for $\dot{H}$ which at the present time is unclear. On the other hand, the existence of the uniqueness condition for the obtained solution is not guaranteed; this mathematical complexity is due to the fact that we are dealing with a cosmological model that is far from equilibrium from the thermodynamics point of view and involves non linear differential equations. As commented previously, the near equilibrium condition is not consistent with an expanding fluid then the non linear approach for bulk viscosity is more appropriated if we intend to describe the late time stage of the universe. Under such conditions we must consider an alternative way to extract cosmological information from the full model. The dynamical system approach has proven to be a powerful tool for studying global dynamical properties of various cosmologies in General Relativity, see for instance \cite{libro} and references therein. The convenience in the use of dynamical systems procedure is that without solving the dynamic equations system completely, one can have qualitative information on important global features of the phase space and the full set of possible solutions, in terms of the critical points of the system and their linear stability. Moreover, in this work the dynamical system technique is useful also to determine that solutions describing accelerated expanding universes exist.

We now proceed to study the linear stability of the cosmological models described in the previous section. As we will see below, we focus on two different cases, one in which dark matter and dark energy do not interact, i.e., $Q=0$ and secondly $Q\neq 0$, which represents an interacting dark sector. Following the line of reasoning of Ref. \cite{lepesys}, we introduce a convenient set of variables to establish the general dynamical system describing the different classes of cosmological models to be investigated, yielding 
\begin{eqnarray}
x&=&\Omega_{DE}=\frac{\rho_{DE}}{3H^2},\\
y&=&\Omega_{m}=\frac{\rho_{m}}{3H^2},\\
z&=&\Omega_{\Pi}=\frac{\Pi}{3H^2}.
\end{eqnarray}
Notice that the above variables are simply the fractional energy density parameters which are dimensionless and bounded, this second feature allows to safely determine a complete set of fixed points. Moreover, having dimensionless dynamical variables is more convenient to manage the numerical calculations. In terms of the aforementioned variables and using the number of e-folds, $N := \ln a$, the set of Eqs. (\ref{eq:transporte}), (\ref{eq:Cont2}) and (\ref{eq:Cont3}), takes the following autonomous form
\begin{eqnarray}
	\frac{dx}{dN}&=&x(3\gamma_{DE}x-3\gamma_{DE}+3z-y-4x+4)-\frac{Q}{3H^3}, \label{eq:dyn_sys1}\\  
	\frac{dy}{dN}&=&3\gamma_{DE}xy +3(y-1)z-y^2+(1-4x)y+\frac{Q}{3H^3}, \\
	\frac{dz}{dN}&=& 3z(\gamma_{DE}x+z)-\frac{3yz}{y+j^2z}\left(\frac{1}{\zeta_0}\sqrt{\frac{y}{3}}+j^2+\frac{y}{z}\right)+z(1-4x-y)-\frac{3z^2}{2y}\nonumber \\ &+&\frac{Q}{6H^3}\frac{z}{y}.\label{eq:dyn_sys3}
\end{eqnarray}
Note that for $j=0$ we recover the dynamical system studied in Ref. \cite{lepesys}, which represents the linear regime of the Israel-Stewart theory; in terms of the variables $(x,y,z)$, the effective parameter state can be written as  
\begin{equation}
	\omega_{\mathrm{eff}}=\gamma_{\mathrm{eff}}-1=\left(\gamma_{DE}-\frac{4}{3}\right)x+z-\frac{y}{3}+\frac{1}{3},
\end{equation}
where we have considered, $\gamma_{r}-1 =\omega_{r} = 1/3$. As can be seen, the above parameter state does not depend explicitly on the parameter $j$, which is the responsible of inducing the non linear bulk viscous effects. It is well known that the critical points of the dynamical system denoted as $P_{i} = (x_{ci},y_{ci},z_{ci})$, (\ref{eq:dyn_sys1})-(\ref{eq:dyn_sys3}) can be obtained by imposing the conditions, $dx/dN = dy/dN = dz/dN = 0$. Therefore the linear stability of the dynamical system around those critical points is studied by computing the eingenvalues associated to the {Jacobian} matrix.

\subsection{Non interacting dark sector}
The critical points of the dynamical system (\ref{eq:dyn_sys1})-(\ref{eq:dyn_sys3}) with $Q=0$ are summarized in table (\ref{tab:pc}). It is worthy to notice that in according to our results, the number of critical points is smaller than those obtained in the linear regime studied in \cite{lepesys}. From the thermodynamics point of view, the critical points of the dynamical system coincide with the stationary thermodynamic states, in this sense we can observe a crucial difference between the linear and non linear regimes of the Israel-Stewart formalism; the non linear description restricts the number of stationary states due to the relaxation of the near equilibrium condition (\ref{eq:near}), physically speaking, the form of the thermodynamic {\it force} and its conjugate thermodynamic {\it flux} (given by $\Pi$), which is the response of the system to reach the equilibrium in the fluid, are different in both schemes. 

Now let us analyze which type of universe is described by each critical solution of our dynamical system in the non interacting regime. Note that the critical point $P_{1}$ describes a radiation dominated universe, this is when $\Omega_{m}= \Omega_{DE} = 0$, which implies $\Omega_{r}=1$ due to the Friedmann constraint. This also can be seen at effective level since $\gamma_{\mathrm{eff}} = 4/3$, then the cosmic evolution is decelerated in this case. 

The critical point $P_{2}$ represents a de Sitter like universe or in other words, a dark energy dominated universe, $\Omega_{DE} = 1$ and $\Omega_{m} = \Omega_{r} = 0$, with $\gamma_{\mathrm{eff}} = \gamma_{DE}$. Thus, solutions close to this fixed point corresponds to a universe under accelerated expansion with $\gamma_{DE} < 2/3$. 

For the critical point $P_{3}$ we obtain a scaled solution as in Ref. \cite{lepesys}. In this case, the following condition is satisfied $\Omega_{DE} =1-z_{c3}/(\gamma_{DE}-1)$, $\Omega_{m} = z_{c3}/(\gamma_{DE}-1)$ and $\Omega_{r} = 0$. This fixed point corresponds to a dark energy dominated solution since $\gamma_{\mathrm{eff}} = \gamma_{DE}$, then the accelerated expansion will take place for values of $\gamma_{DE}$ given in the aforementioned interval. This scaled solution turns out to be interesting if we appeal to the definition of the coincidence parameter, $r$, namely

\begin{equation}
    r := \frac{\rho_{m}}{\rho_{DE}} = \frac{y}{x} = \frac{z_{c3}}{\gamma_{DE}-1-z_{c3}}.
\end{equation}

According to the intervals established for the existence of the critical point $P_{3}$,  {the above ratio of densities} takes values around the unity, and consequently, this critical point represents an universe in which the cosmological coincidence problem could be alleviated.

The stability of each point is established on the last column of the table (from left to right). In this case we define

\begin{equation}
\zeta_0^{*}=\frac{2(\gamma_{DE}-1)}{\sqrt{3}(\gamma_{DE}^2-2\gamma_{DE}-1)((\gamma_{DE}-1)j^2+1)},
\end{equation}

which corresponds to the viscosity coefficient for the critical solution corresponding to $P_{3}$ for a dark energy dominated universe. On the other hand, we also defined $z_{c3} := (\zeta_{0}/\zeta_0^{*})^{2}$. As can be seen, the stability criteria for $P_{3}$ depends on the $j$ parameter.\\
\begin{table}
	\begin{tabular}{c c c c c c c}
			\hline \hline\\
			\makebox[2cm][c]{Point $P_i$} & \makebox[2cm][c]{$x_{ci}$} & \makebox[2cm][c]{$y_{ci}$} & \makebox[2cm][c]{$z_{ci}$} & \makebox[2cm][c]{Existence} & \makebox[2cm][c]{$\gamma_{\mathrm{eff}}$} & \makebox[2cm][c]{Stability} \\
			\hline \\
			\makebox[2cm][c]{$P_1$} & \makebox[2cm][c]{0} & \makebox[2cm][c]{0} & \makebox[2cm][c]{0} & \makebox[2cm][c]{Always} & \makebox[2cm][c]{$\frac{4}{3}$} & \makebox[2cm][c]{Saddle Point}\\
			\\
			\makebox[2cm][c]{$P_2$} & \makebox[2cm][c]{1} & \makebox[2cm][c]{0} & \makebox[2cm][c]{0} & \makebox[2cm][c]{Always} & \makebox[2cm][c]{$\gamma_{DE}$} & \makebox[2cm][c]{Stable spiral}\\
			\\
			\makebox[2cm][c]{} & \makebox[2cm][c]{} & \makebox[2cm][c]{} & \makebox[2cm][c]{} & \makebox[2cm][c]{} & \makebox[2cm][c]{} & \makebox[2cm][c]{for $\gamma_{DE}<0$.}\\
			\\
			\makebox[2cm][c]{} & \makebox[2cm][c]{} & \makebox[2cm][c]{} & \makebox[2cm][c]{} & \makebox[2cm][c]{} & \makebox[2cm][c]{} & \makebox[2cm][c]{Saddle point for}\\
			\\
			\makebox[2cm][c]{} & \makebox[2cm][c]{} & \makebox[2cm][c]{} & \makebox[2cm][c]{} & \makebox[2cm][c]{} & \makebox[2cm][c]{} & \makebox[2cm][c]{$0<\gamma_{DE}<\frac{4}{3}$ and}\\
			\\
			\makebox[2cm][c]{} & \makebox[2cm][c]{} & \makebox[2cm][c]{} & \makebox[2cm][c]{} & \makebox[2cm][c]{} & \makebox[2cm][c]{} & \makebox[2cm][c]{$\frac{4}{3}<\gamma_{DE}<2$.}\\
			\\
			\makebox[2cm][c]{$P_3$} & \makebox[2cm][c]{$1-\frac{z_{c3}}{\gamma_{DE}-1}$} & 	\makebox[2cm][c]{$\frac{z_{c3}}{\gamma_{DE}-1}$} & 	\makebox[2cm][c]{$z_{c3}$} & 	\makebox[2cm][c]{For} & \makebox[2cm][c]{$\gamma_{DE}$} & \makebox[2cm][c]{Stable Node for}\\
			\\
			\makebox[2cm][c]{} & \makebox[2cm][c]{} & \makebox[2cm][c]{} & \makebox[2cm][c]{} & \makebox[2cm][c]{$\gamma_{DE}-1<z_{c3}<0$} & \makebox[2cm][c]{} & \makebox[2cm][c]{$\zeta_0<\zeta_0^{*}$ and}\\
			\\
			\makebox[2cm][c]{} & \makebox[2cm][c]{} & \makebox[2cm][c]{} & \makebox[2cm][c]{} & \makebox[2cm][c]{and $1-\sqrt{2}<\gamma_{DE}<1$} & \makebox[2cm][c]{} & \makebox[2cm][c]{$1-\sqrt{2}<\gamma_{DE}<1$.}\\
			\\
			\makebox[2cm][c]{} & \makebox[2cm][c]{} & \makebox[2cm][c]{} & \makebox[2cm][c]{} & \makebox[2cm][c]{} & \makebox[2cm][c]{} & \makebox[2cm][c]{Saddle point for}\\
			\\
			\makebox[2cm][c]{} & \makebox[2cm][c]{} & \makebox[2cm][c]{} & \makebox[2cm][c]{} & \makebox[2cm][c]{} & \makebox[2cm][c]{} & \makebox[2cm][c]{$\zeta_0>\zeta_0^{*}$ and}\\
			\\
			\makebox[2cm][c]{} & \makebox[2cm][c]{} & \makebox[2cm][c]{} & \makebox[2cm][c]{} & \makebox[2cm][c]{} & \makebox[2cm][c]{} & \makebox[2cm][c]{$1-\sqrt{2}<\gamma_{DE}<1$.}\\
			\\
			\hline \hline
				\end{tabular}
		\caption{Phase space coordinates of the fixed points of the dynamical system in the non interacting regime ($Q=0$). The fifth column (from left to right) shows the respective ranges of parameters where the fixed points do exist. The next columns show the values of the effective parameter state, $\gamma_{\mathrm{eff}}$ and the linear stability of the fixed point according to the Liapounov classification.}
		\label{tab:pc}
\end{table}
			
\subsection{Interacting dark sector}
For an interacting dark sector we will consider the following interaction term
\begin{equation}
	Q = \frac{\lambda \rho_m \rho_{DE}}{H} = 9\lambda H^3 x y, \label{eq:qterm}
\end{equation}
in the dynamical system (\ref{eq:dyn_sys1})-(\ref{eq:dyn_sys3}). $\lambda$ corresponds to a coupling parameter that controls the interaction between dark matter and dark energy. Notice that $Q$ is no longer a free function and rather it depends on the dynamics of the cosmological model. It is worthy to mention that the form of the $Q$-term is generally taken to be proportional to the energy densities of the dark sector, this generic election provides in most of the cases a solution to the cosmological coincidence problem \cite{qcoinci1, qcoinci2}. In the context of the standard cosmological model, a well known fact is the appearance of increasing discrepancies, or tensions, between some predictions such as $H_0$ and $\sigma_8$ (to mention some) according to different independent observational datasets. Specifically, there exists a tension between estimations of the present expansion rate of the universe quantified by $H_0$ corresponding to early universe observations (such as those of CMB anisotropies made by the Planck collaboration \cite{planck}) and late universe ones (such as distances of SNaI or LSS surveys data such as DES, KIDS, etc \cite{des, kids}). Similarly,  estimations of  $\sigma_8$ characterizing the amount of matter clustering at late times inferred from early and late time datasets are in tension. This problem represents a challenge in modern cosmology, if such discrepancies can not be attributed to systematic errors in the data then we have clear evidence of the failure of the $\Lambda$CDM model; this opens the door to alternative cosmological models. In Ref. \cite{valentino} can be found a complete compilation of alternative cosmological models which are viable to provide a solution for the $H_{0}$ tension, we would like to comment that in the aforementioned reference the interacting dark sector and bulk viscous models appear as good candidates in this contest. Furthermore, in \cite{dmde} it is shown that by allowing the interaction in the dark sector the $H_{0}$ tension is alleviated with the consideration of an interaction term proportional to $\rho_{DE}$. In addition, in Ref. \cite{halos}, interacting dark energy is considered with $Q \propto \rho_{DE}$ in order to study large structure formation. In the literature we can find a wide host of functional forms of the $Q$-term \cite{wang}. It is quite common to adopt $Q \propto \rho_{m}$ or $Q \propto \rho_{DE}$, however, within these schemes, the interaction term is not sensitive to changes of the not elected energy density. A viable interesting description from the physical point of view for the interacting framework, is given by a $Q$-term written as the product of both energy densities; in this case any change in the energy densities will affect the $Q$-term. Besides, in Ref. \cite{lotka1, lotka2, lotka3} it can be found that an interaction term of this type allows to write the dynamical equations of the interacting dark sector as a set of Lotka-Volterra-like equations which lead to a cyclic interchange of energy between those components, providing a more realistic description for the universe.\\

In order to carry out the asymptotic stability analysis, we firstly determine the fixed points of the dynamical system with interaction term (\ref{eq:qterm}). As in the non interacting dark sector case, we obtain again less critical points than in the linear regime studied in \cite{lepesys} for interacting dark sector, we can also observe differences between the critical points obtained here and those of the analysis given in \cite{lepesys} since the form for interaction term is distinct. The coordinates in the phase space for the critical points are given by $P^Q_{1}=(0,0,0)$, $P^Q_{2}=(1,0,0)$ and $P^Q_{3}=\Big(-\frac{1}{3 \lambda},\frac{4-3\lambda}{3\lambda},0\Big)$.

It is important to stress out that the phase space coordinates of these fixed points do not depend on the $j$ parameter, hence their existence and position in the phase space is independent on whether the theory lies in the linear or non linear regimes.

$P^{(Q)}_{1}$ corresponding to a radiation dominated universe, is independent of $\lambda$, that is, of the strength of the interaction between dark matter and dark energy. This point is classified as a saddle point and the eigenvalues of the Jacobian matrix are $4$, $-2$ and $4-3\gamma_{DE}$. The fixed point $P^{(Q)}_{2}$ again describes a de Sitter universe. This stationary solution turns to be quite interesting in physical terms, linearized solutions around this point are able to describe our present universe. The eigenvalues of the Jacobian matrix in this case are, $3\gamma_{DE}-4,\frac{3}{4}(4\gamma_{DE}-\sqrt{\lambda^2+16}+3\lambda-4)$ and $\frac{3}{4}(4\gamma_{DE}+\sqrt{\lambda^2+16}+3\lambda-4)$, we will give more details about this point below. The last fixed point seems to be a scaled solution but it lacks of physical meaning, then we discard it from our analysis. 

In order to finish this section, let us to comment that for the non interacting case, the physical information extracted from the critical points obtained, is similar to the linear regime explored in \cite{lepesys}. However, for the interacting dark sector we obtain remarkable differences between both scenarios from the physical point of view.

\subsubsection{Linear stability analysis for \texorpdfstring{$P^{(Q)}_{2}$}{}}
In order to study solutions of the dynamical system that possibly describe the expansion of the universe at late stages of its evolution, we focus our analysis to the de Sitter like critical point since linearized solutions around it  describe an accelerated expanding universe.

\begin{figure}[htbp!]
\centering
\includegraphics[scale=0.6]{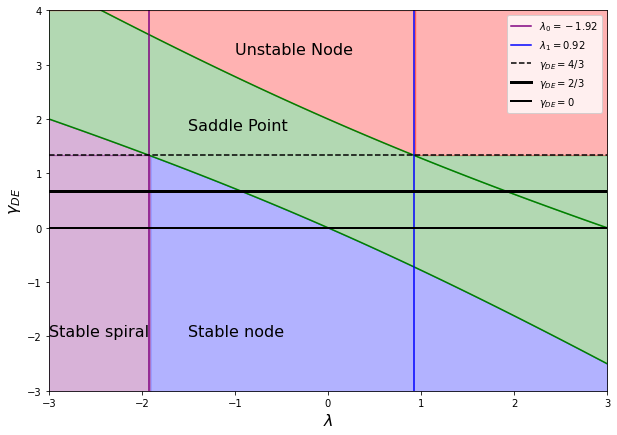}
\caption{Stability of the critical point $P^Q_{2}$ in the space of parameters $(\lambda, \gamma_{DE})$.}
\label{fig:Parameter_Space_C1_P2}
\end{figure}

The linear stability of each critical point is established according to the Liapunov criteria, depending on the sign of the eigenvalues of the Jacobian matrix. Since the eigenvalues only depend on $\gamma_{DE}$ and $\lambda$, in Figure \ref{fig:Parameter_Space_C1_P2} we illustrate the stability behavior of $P^Q_{2}$ in the parameter space $(\lambda, \gamma_{DE})$. Appendix \ref{app:stabi} is devoted to the conditions that must be satisfied by the parameters of the model for each case. As can be seen in the plot, the dashed and the solid lines represent some cases of interest in the fluid description. The dashed line is for radiation, for the case $\gamma_{DE} = 2/3$, we enter in the quintessence region and for $\gamma_{DE} = 0$ we have a cosmological constant like fluid. The negative region of the vertical axis is associated to a phantom like behavior, it is worthy to mention that depending on the value of the $\lambda$ parameter, the stability of the critical point can be guaranteed within the phantom region; the stability conditions for this critical point do not depend on the $\zeta_{0}$ parameter.

\section{Analytical linear solutions in de Sitter universe}
\label{sec:linearized}
Now we study the behavior of linear solutions around the critical points $P_{2}$ and $P^Q_{2}$, which represent an universe dominated by dark energy in the non interacting and interacting dark sector, respectively. The linear dynamical system is given by
\begin{equation}
    \frac{d \vec{x}}{dN}=J \vec{x},
\end{equation}
where $J$ is the Jacobian matrix and $\vec{x} = (x,y,z)$. For the cases of interest we have
\begin{eqnarray}
	\frac{dx}{dN}&=&(3\gamma_{DE}-4)x-(3\lambda+1)y+3z,  \label{eq:lin1}\\ 
	\frac{dy}{dN}&=&3(\gamma_{DE}+\lambda-1)y-3z, \\ 
	\frac{dz}{dN}&=&-3y+3\left(\gamma_{DE}+\frac{\lambda}{2}-1\right)z, \label{eq:lin3}
\end{eqnarray}
where we have considered the interaction term given in (\ref{eq:qterm}). Therefore $\lambda=0$ stands for the non interacting dark sector case. By solving the previous dynamical equations with no interaction term, we obtain
\begin{eqnarray}
	x(N)&=&-\frac{1}{2} e^{-6N} \left[-2 e^{2 N}(x_0+y_0)+e^{6 N}(y_0-z_0)+y_0+z_0\right], \label{eq:linq01}\\ 
	y(N)&=&\frac{1}{2} e^{-6N} \left(y_0 e^{6 N}-z_0 e^{6 N}+y_0+z_0\right), \\ z(N)&=&\frac{1}{2} e^{-6N} \left(-y_0 e^{6 N}+z_0 e^{6 N}+y_0+z_0\right)\label{eq:linq03},
\end{eqnarray}
being $x_{0}, y_{0}, z_{0}$ integration constants corresponding to the initial condition, $N=0$. If we now solve for the interacting case ($\lambda \neq 0$), one gets
\begin{eqnarray}
	x(N)&=& \Theta(N)\left\lbrace 2 \lambda_0^2 e^{\frac{1}{4} \left(3 \lambda_0-9 \lambda -4\right) N} (x_0+y_0)+e^{\frac{3}{2} \lambda_0 N} \left[4 \lambda_0 z_0-\left(\lambda ^2+\lambda_0 \lambda +16\right) y_0\right] + \right. \nonumber \\  
	&+& \left. \left(-\lambda ^2+\lambda_0 \lambda -16\right) y_0 - 4 \lambda_0 z_0 \right\rbrace,  \label{eq:linq11}\\
	y(N)&=& \Theta(N)\left\lbrace y_0 \left[\lambda ^2-\lambda_0 \lambda +\left(\lambda ^2+\lambda_0 \lambda +16\right) e^{\frac{3}{2} \lambda_0 N}+16\right]-4 \lambda_0 z_0 \left(e^{\frac{3}{2} \lambda_0 N}-1\right)\right\rbrace, \\ 
	z(N)&=&\Theta(N) \left\lbrace z_0 \left[\lambda ^2+\lambda_0 \lambda +\left(\lambda ^2-\lambda_0 \lambda +16\right) e^{\frac{3}{2} \lambda_0 N}+16\right]-4 \lambda_0 y_0 \left(e^{\frac{3}{2} \lambda_0 N}-1\right)\right\rbrace,  \label{eq:linq13}
\end{eqnarray}
where we have defined the quantities $\lambda_0 := \sqrt{\lambda^2+16}$ and $\Theta(N):=\frac{e^{-\frac{3}{4}\left(\lambda_0-3 \lambda +4\right) N}}{2\lambda_0^2}$, for simplicity in the notation. Note that for our previous solutions we have considered the special case, $\gamma_{DE}=0$, which represents a cosmological constant like fluid. 

\subsection{{Comparison between cosmological parameters derived from analytical linear solutions and exact numerical solutions }}
\label{sec:theor}
\begin{figure}[htbp!]
\includegraphics[scale=0.36]{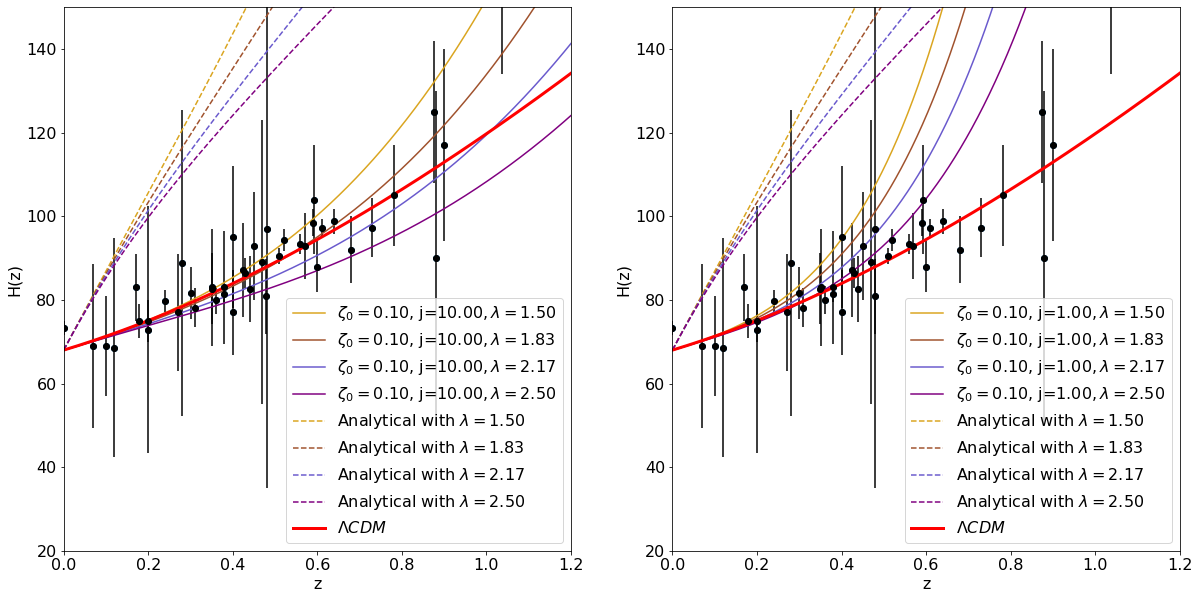}
\caption{The Hubble parameter and observational data from Ref. \cite{magana, mukherjee}.}
\label{fig:H_data}
\end{figure}

In terms of the e-folds number, $N$, the acceleration equation (\ref{eq:Friedmann2}) can be penned as follows
\begin{eqnarray}
    \frac{d}{dN}(\ln H) +1 &=& -\frac{1}{2} \left(\frac{1}{3H^2}\right)(\rho+3p_{\mathrm{eff}})=-\frac{1}{2} \left[\Omega_{DE}+\Omega_m+3(-\Omega_{DE}+\Omega_\Pi)\right],\nonumber \\
    &=& x(N) - \frac{1}{2}y(N) - \frac{3}{2}z(N),\label{eq:int}
\end{eqnarray}
where the following values, $\Omega_{r} = \gamma_{DE} = 0$, were taken into account. Then, the integration of the above expression will give us the explicit form of the Hubble parameter as, $H=H(N)$. Note that the resulting expression for $H$ will carry the information of the variables $x,y,z$ of the dynamical system analyzed in the previous sections. We can perform the numerical integration of Eq. (\ref{eq:int}) by considering different values for the set of parameters of the model. For the deceleration parameter, $q$, can be found a similar expression as the one above, by considering the standard definition for this parameter, $1+q := -\dot{H}/H^{2}$, one gets
\begin{equation}
    q(N) = \frac{1}{2} \left(\Omega_{DE}+\Omega_m+3(-\Omega_{DE}+\Omega_\Pi)\right) = \frac{1}{2}y(N)+\frac{3}{2}z(N)-x(N).
\end{equation}
In Fig. \ref{fig:H_data} we show the behavior of the Hubble parameter obtained from the numerical integration given in Eq. (\ref{eq:int}) and we compare it with the Hubble parameter of the $\Lambda$CDM model. We focus on the case where the interaction in the dark sector is allowed. In both panels the dashed lines were obtained by inserting the analytical linear solutions (\ref{eq:linq11}) - (\ref{eq:linq13}) in Eq. (\ref{eq:int}). On the other hand, the solid lines also come from the integration of Eq. (\ref{eq:int}) but in this case the $x,y,z$ variables are obtained from the numerical solution of the dynamical system (\ref{eq:dyn_sys1}) - (\ref{eq:dyn_sys3}) with $\gamma_{DE} = 0$ and the interaction term (\ref{eq:qterm}). These latter solutions depend explicitly on the parameters $j$ and $\zeta_{0}$. According to the behavior depicted in the plots we can clearly observe that the Hubble parameter constructed with the analytical linear solutions does not fit properly the observational data and given that $\gamma_{DE} = 0$; for comparison reasons with $\Lambda$CDM model, then the parameter $\lambda$ plays a relevant role to determine its behavior; as this parameter takes different values, we do not observe alterations in the nature of $H$. However, the non linear regime of the Israel-Stewart model offers a better theoretical framework since the resulting Hubble parameter can fit the data and for low redshift values is practically indistinguishable from the corresponding one to $\Lambda$CDM model. Note that in this case the solutions obtained from the dynamical system (\ref{eq:dyn_sys1}) - (\ref{eq:dyn_sys3}) are very sensitive to changes of the $j$ parameter, which characterizes the non linear effects; as the non linear effects become stronger, the resulting Hubble parameter is closer to the $\Lambda$CDM model and data points, otherwise the model departs from the observational data points and $\Lambda$CDM model as the redshift values increases. We must emphasize that in the non linear regime the Hubble parameter exhibits a decreasing behavior (from past to present time given at $z=0$) as the cosmological standard model; this feature is relevant to describe properly the cosmic evolution of the observable universe. 
\\

\begin{figure}[htbp!]
\includegraphics[scale=0.31]{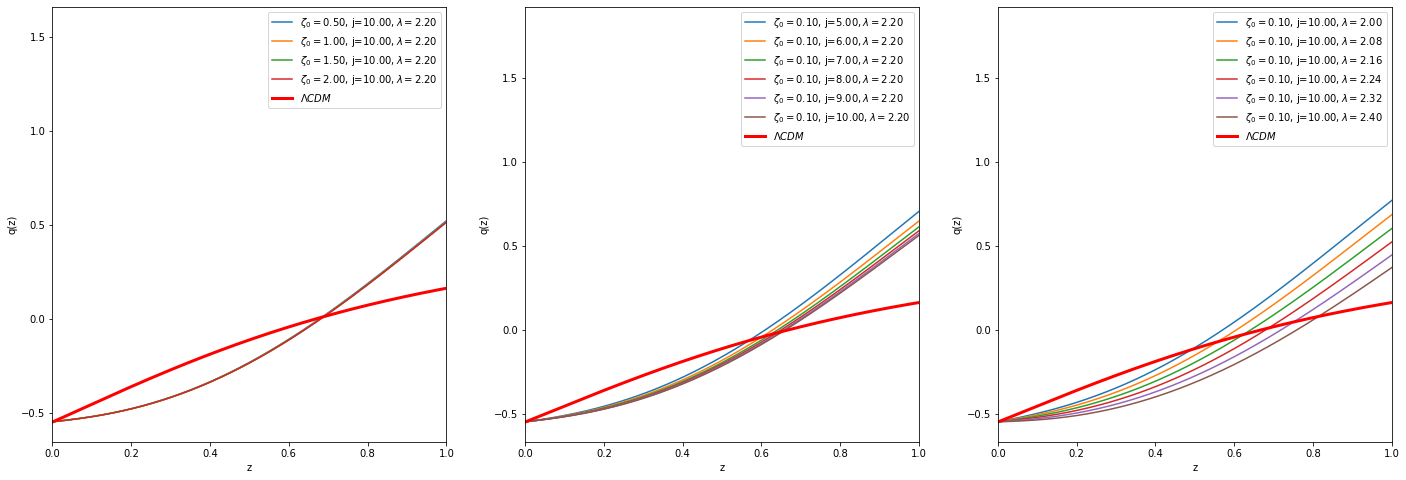}
\caption{Deceleration parameter as a function of the redshift.}
\label{fig:decel}
\end{figure}

Based on the behavior exhibited by the Hubble parameter for the analytical linear solutions and the full non linear regime in Fig. \ref{fig:H_data}, we now focus only on the deceleration parameter constructed with the $x,y,z$ variables obtained from the dynamical system (\ref{eq:dyn_sys1}) - (\ref{eq:dyn_sys3}) for the interacting dark sector case, as commented previously and compare it with the $\Lambda$CDM model expression given as
\begin{equation}
    q(z) = -1 + \frac{3}{2\left[1+\frac{\Omega_{\Lambda,0}}{\Omega_{m,0}}(1+z)^{-3}\right]}.
\end{equation}
We show the deceleration parameter as a function of the redshift $z$ in Fig. \ref{fig:decel} by considering different sets of values for the parameters of the model, $\zeta_{0}$, $j$ and $\lambda$. As can be seen in the first panel (from left to right), for fixed values of $j$ and $\lambda$, the deceleration parameter is identical for different values of $\zeta_{0}$, which characterizes the viscosity. The middle panel shows that slightly variations on the curves of the deceleration parameter can be obtained if one considers different values of  $j$ and keeps fixed values for the parameters $\zeta_{0}$ and $\lambda$. For the third panel (from left to right) we observe that each case can be clearly distinguished if we consider different values for $\lambda$ and keep fixed values for $\zeta_{0}$ and $j$. A remarkable fact about the cases shown in Fig. \ref{fig:decel} is that the deceleration parameter always has a transition from positive to negative values for a specific value of redshift, $z_{t}$, as the $\Lambda$CDM model; this behavior is independent of the values assigned to the parameters of the model. This means that the model has a transition from a decelerated expansion to an accelerated one. Note that in all cases we have $z_{t} > 0$, i.e., the transition takes place at the past. This is consistent with the recent accelerated expansion of our universe that began around $z \sim 0.6$, the transition from decelerated to accelerated expansion is independent of the cosmological model, see Ref. \cite{recent}. Besides, at present time, $z=0$, our results coincide with the $\Lambda$CDM model. We would also like to comment that the value of $z_{t}$ is determined by the election of the values for the parameters $\zeta_{0}$, $j$ and $\lambda$.            

\section{Concluding remarks}
\label{sec:conclusions}
In this work we have considered the inclusion of dissipative effects in the matter sector of the cosmic fluid description; the approach for such effects is given by a non linear extension of the Israel-Stewart theory. The interacting scenario between dark matter and dark energy was allowed in our scheme. We adopted the dynamical system perspective in order to provide a global description of the model and we also compared our results with those obtained in the linear regime of the Israel-Stewart theory, reported in Ref. \cite{lepesys}. The first difference between the non linear and linear regime of the Israel-Stewart model is the number of critical points allowed by the dynamical system, the number of points in the non linear scheme is less. However, the critical points obtained in this work for the non interacting dark sector preserve some features of the critical points found in the linear regime for $Q=0$. By turning on the interaction term for the dark sector we observe that the physical information obtained from the critical points in the non linear approach differs from the one obtained in the linear regime. In our case the critical point of interest is $P_{2}^{Q}$ and represents an universe dominated by dark energy, a point of this kind also appears for the non interacting dark sector and we denote it as $P_{2}$. The stability criteria of the critical point do not depend on the parameters that characterize the non linear and viscous effects, $j$ and $\zeta_{0}$. Instead, the stability for such point can be established by considering a reduced space of parameters given by $(\lambda, \gamma_{DE})$, i.e., the coupling constant appearing in the interaction $Q$-term and the parameter state of the dark energy sector. For the non interacting case we observe that the stability of $P_{2}$ is determined only by $\gamma_{DE}$. As shown previously, for quintessence, cosmological constant and phantom like fluids, $P_{2}^{Q}$ exhibits regions of stability in the parameter space.\\

We consider the numerical solution coming from the full dynamical system and the analytical solutions coming from the set of linear differential equations. For the couple of solutions we also consider the special case, $\gamma_{DE} = 0$, since we compare the corresponding expansion rate with the one obtained in $\Lambda$CDM model. As discussed in the text, the full non linear regime approach is more favored by observational data than analytical linear solutions; we observe that the $j$ parameter plays the role of a tuner between the expansion rate and data points; as the non linear effects increase (characterized by $j$), better fit of observations we obtain. The sensitivity of the Hubble parameter with respect to changes on the values of $\zeta_{0}$ and $\lambda$ is low. It is also notable that for low redshift values, the Hubble rates of the full non linear regime and $\Lambda$CDM model are identical, this indicates that the inclusion of viscous effects in the matter sector described by the non linear regime of the Israel-Stewart theory can be a viable theoretical framework to explain the recent accelerated expansion of our universe. We leave for future investigation the viability analysis of this viscous interacting model as solver of the $H_{0}$ tension.\\ 

Finally, we observe that the deceleration parameter constructed with the numerical solutions obtained from the full dynamical system have a desirable behavior, at the past is positive and for a specific value of the redshift, $z_{t} > 0$, becomes negative. A transition from decelerated to accelerated cosmic expansion is unveiled in this cosmological model. As far as we know, this behavior corresponds to an important characteristic of the observable universe. The transition value $z_{t}$ is directly related with the values of the parameters $\lambda$, $\zeta_{0}$ and $j$.

\appendix
\section{Classification of linear stability conditions for \texorpdfstring{$P^Q_{2}$}{}}
\label{app:stabi}
For simplicity in our notation, let us define the following constants
\begin{align}
& \lambda_0=\frac{1}{6}(-3-\sqrt{73}), \\
& \lambda_1=\frac{1}{6}(\sqrt{73}-3), \\
& \gamma_m=\frac{1}{4}(4-3\lambda)-\frac{\sqrt{\lambda^2+16}}{4},\\
& \gamma_M=\frac{\sqrt{\lambda^2+16}}{4}+ \frac{1}{4}(4-3\lambda).
\end{align}
\begin{enumerate}
    \item $P^Q_{2}$ is a {\bf stable node} if the parameters lie in the subset of the parameter space given by $C_1\cup C_2$ where 
\begin{align}
& C_1=\left\lbrace(\lambda,\gamma_{DE}):\lambda \leq \lambda_0 \hspace{0.5cm} \mathrm{and} \hspace{0.5cm} \gamma_{DE}<\frac{4}{3}\right\rbrace, \\
& C_2=\left\lbrace(\lambda,\gamma_{DE}):\lambda \geq \lambda_0 \hspace{0.5cm} \mathrm{and} \hspace{0.5cm} \gamma_{DE}<\gamma_m(\lambda)\right\rbrace.
\end{align}
\item $P^Q_{2}$ is a {\bf saddle point} if the parameters lie inside the subset of the parameter space given by $C_1\cup C_2 \cup C_3 \cup C_4 \cup C_5$ where 
\begin{align}
& C_1=\left\lbrace(\lambda,\gamma_{DE}):\lambda_0<\lambda_{DE} \leq\lambda_1 \hspace{0.5cm} \mathrm{and} \hspace{0.5cm} \gamma_m(\lambda)<\gamma_{DE}<\frac{4}{3}\right\rbrace ,\\
& C_2=\left\lbrace(\lambda,\gamma_{DE}):\lambda >\lambda_1 \hspace{0.5cm} \mathrm{and} \hspace{0.5cm} \gamma_m(\lambda)<\gamma_{DE}<\gamma_M(\lambda)\right\rbrace,\\
& C_3=\left\lbrace(\lambda,\gamma_{DE}):\lambda > \lambda_1 \hspace{0.5cm} \mathrm{and} \hspace{0.5cm} \gamma_M(\lambda)<\gamma_{DE}<\frac{4}{3}\right\rbrace,\\
& C_4=\left\lbrace(\lambda,\gamma_{DE}):\lambda \leq \lambda_0 \hspace{0.5cm} \mathrm{and} \hspace{0.5cm} \gamma_m(\lambda)<\gamma_{DE}<\gamma_M(\lambda)\right\rbrace,\\
& 	C_5=\left\lbrace(\lambda,\gamma_{DE}):\lambda_0<\lambda <\lambda_1 \hspace{0.5cm} \mathrm{and} \hspace{0.5cm} \frac{4}{3}<\gamma_{DE}<\gamma_M(\lambda)\right\rbrace.
\end{align}
\item $P^Q_{2}$ is {\bf unstable node} if the parameters lie inside the subset $C_1\cup C_2$ of the parameter space where
\begin{align}
& C_1=\left\lbrace(\lambda,\gamma_{DE}):\lambda \leq \lambda_0 \hspace{0.5cm} \mathrm{and} \hspace{0.5cm} \gamma_{DE}>\gamma_M(\lambda)\right\rbrace,\\
& C_2=\left\lbrace(\lambda,\gamma_{DE}):\lambda > \lambda_0 \hspace{0.5cm} \mathrm{and} \hspace{0.5cm} \gamma_{DE} > \frac{4}{3}\right\rbrace.
\end{align}
\item $P^Q_{2}$ is {\bf stable spiral} if the parameters lie inside $C_1\cup C_2 \cup C_3$ where 
\begin{align}
& C_1=(\lambda =\lambda_0)\cap (\gamma_{DE} <\gamma_3),\\
& C_2= (\lambda_0<\lambda <\lambda_1) \cap  ((\gamma_{DE} <\gamma_m)\cup(
  \frac{4}{3}<\gamma_{DE} <\gamma_M)),\\
&  C_3=A\cup B,
\end{align}
with the following definitions
\begin{equation}
    A=\{\lambda =\lambda_1 \,\, y \,\, \gamma <\gamma_s\}, \ \ \ \ \ B=C\cap D, 
\end{equation}
and 
\begin{equation}
 C=\{\lambda >\lambda_1\}, \ \ \ \ \ D=\{\gamma_{DE} <\gamma_m\}\cup \left\lbrace \gamma_M<\gamma_{DE} <\frac{4}{3}\right\rbrace.    
\end{equation}
\end{enumerate}

\acknowledgments
G. A. P. acknowledges CONACyT doctoral grant. This work has been supported by S.N.I. CONACyT-M\'exico (M. C. and A. A. L.)


\bibliographystyle{ieeetr}
\bibliography{bibliography}

\begin{thebibliography}{10}

\bibitem{growth1}
L.~Espinosa-Portales and J.~Garcia-Bellido, ``{Covariant formulation of
  non-equilibrium thermodynamics in General Relativity},'' {\em Phys. Dark
  Univ.}, vol.~34, p.~100893, 2021.

\bibitem{growth2}
J.~Garcia-Bellido and L.~Espinosa-Portales, ``{Cosmic acceleration from first
  principles},'' {\em Phys. Dark Univ.}, vol.~34, p.~100892, 2021.

\bibitem{entro}
W.~Zimdahl, D.~Pav\'on, and J.~Triginer, ``{Cosmology with bulk pressure},''
  {\em Helv. Phys. Acta}, vol.~69, p.~225, 1996.

\bibitem{data1}
A.~G. Riess {\em et~al.}, ``{Observational evidence from supernovae for an
  accelerating universe and a cosmological constant},'' {\em Astron. J.},
  vol.~116, p.~1009, 1998.

\bibitem{data2}
S.~Perlmutter {\em et~al.}, ``{Measurements of $\Omega$ and $\Lambda$ from 42
  high redshift supernovae},'' {\em Astrophys. J.}, vol.~517, p.~565, 1999.

\bibitem{data3}
C.~L. Bennett {\em et~al.}, ``{Nine-Year Wilkinson Microwave Anisotropy Probe
  (WMAP) Observations: Final Maps and Results},'' {\em Astrophys. J. Suppl.},
  vol.~208, p.~20, 2013.

\bibitem{data4}
N.~Suzuki {\em et~al.}, ``{The Hubble Space Telescope Cluster Supernova Survey:
  V. Improving the Dark Energy Constraints Above z\ensuremath{>}1 and Building
  an Early-Type-Hosted Supernova Sample},'' {\em Astrophys. J.}, vol.~746,
  p.~85, 2012.

\bibitem{origin}
W.~Zimdahl, ``{'Understanding' cosmological bulk viscosity},'' {\em Mon. Not.
  Roy. Astron. Soc.}, vol.~280, p.~1239, 1996.

\bibitem{eck}
C.~Eckart, ``The thermodynamics of irreversible processes. iii. relativistic
  theory of the simple fluid,'' {\em Phys. Rev.}, vol.~58, p.~919, 1940.

\bibitem{is}
W.~Israel and J.~M. Stewart, ``{Transient relativistic thermodynamics and
  kinetic theory},'' {\em Annals Phys.}, vol.~118, p.~341, 1979.

\bibitem{obs1}
A.~Hern\'andez-Almada, ``{Cosmological test on viscous bulk models using Hubble
  Parameter measurements and type Ia Supernovae data},'' {\em Eur. Phys. J. C},
  vol.~79, no.~9, p.~751, 2019.

\bibitem{obs2}
N.~Cruz, E.~Gonz\'alez, and G.~Palma, ``{Testing dissipative dark matter in
  causal thermodynamics},'' {\em Mod. Phys. Lett. A}, vol.~36, no.~06,
  p.~2150032, 2021.

\bibitem{obs3}
N.~D.~J. Mohan, A.~Sasidharan, and T.~K. Mathew, ``{Bulk viscous matter and
  recent acceleration of the universe based on causal viscous theory},'' {\em
  Eur. Phys. J. C}, vol.~77, no.~12, p.~849, 2017.

\bibitem{obs4}
N.~Cruz, E.~Gonz\'alez, S.~Lepe, and D.~S\'aez-Chill\'on~G\'omez, ``{Analysing
  dissipative effects in the $\Lambda$CDM model},'' {\em JCAP}, vol.~12,
  p.~017, 2018.

\bibitem{obs5}
N.~Cruz, E.~Gonz\'alez, and G.~Palma, ``{Exact analytical solution for an
  Israel\textendash{}Stewart cosmology},'' {\em Gen. Rel. Grav.}, vol.~52,
  no.~6, p.~62, 2020.

\bibitem{obs6}
W.~Yang, S.~Pan, E.~Di~Valentino, A.~Paliathanasis, and J.~Lu, ``{Challenging
  bulk viscous unified scenarios with cosmological observations},'' {\em Phys.
  Rev. D}, vol.~100, no.~10, p.~103518, 2019.

\bibitem{obs7}
A.~Avelino and U.~Nucamendi, ``{Can a matter-dominated model with constant bulk
  viscosity drive the accelerated expansion of the universe?},'' {\em JCAP},
  vol.~04, p.~006, 2009.

\bibitem{large}
H.~Velten, T.~R.~P. Caram\^es, J.~C. Fabris, L.~Casarini, and R.~C. Batista,
  ``{Structure formation in a $\Lambda$ viscous CDM universe},'' {\em Phys.
  Rev. D}, vol.~90, no.~12, p.~123526, 2014.

\bibitem{cqg95}
R.~Maartens, ``{Dissipative cosmology},'' {\em Class. Quant. Grav.}, vol.~12,
  p.~1455, 1995.

\bibitem{maartens}
R.~Maartens, ``{Causal thermodynamics in relativity},'' 9 1996.

\bibitem{mendez}
R.~Maartens and V.~Mendez, ``{Nonlinear bulk viscosity and inflation},'' {\em
  Phys. Rev. D}, vol.~55, pp.~1937--1942, 1997.

\bibitem{causal}
F.~S. Bemfica, M.~M. Disconzi, and J.~Noronha, ``{Causality of the
  Einstein-Israel-Stewart Theory with Bulk Viscosity},'' {\em Phys. Rev.
  Lett.}, vol.~122, no.~22, p.~221602, 2019.

\bibitem{Coley}
A.~A. Coley, {\em {Dynamical systems and cosmology}}.
\newblock Dordrecht, Netherlands: Kluwer, 2003.

\bibitem{BoyceDiPrima}
W.~E. Boyce and R.~C.~Di~Prima, {\em {Elementary differential equations and
  boundary value problems}}.
\newblock New York: John Wiley, 1969.

\bibitem{lepesys}
S.~Lepe, G.~Otalora, and J.~Saavedra, ``{Dynamics of viscous cosmologies in the
  full Israel-Stewart formalism},'' {\em Phys. Rev. D}, vol.~96, no.~2,
  p.~023536, 2017.

\bibitem{yoelsy}
A.~Avelino, Y.~Leyva, and L.~A. Ure\~na L\'opez, ``{Interacting viscous dark
  fluids},'' {\em Phys. Rev. D}, vol.~88, p.~123004, 2013.

\bibitem{qual1}
A.~A. Coley, R.~J. van~den Hoogen, and R.~Maartens, ``{Qualitative viscous
  cosmology},'' {\em Phys. Rev. D}, vol.~54, p.~1393, 1996.

\bibitem{qual2}
V.~Mendez and J.~Triginer, ``{Qualitative analysis of causal cosmological
  models},'' {\em J. Math. Phys.}, vol.~37, p.~2906, 1996.

\bibitem{qual3}
A.~Sasidharan and T.~K. Mathew, ``{Phase space analysis of bulk viscous matter
  dominated universe},'' {\em JHEP}, vol.~06, p.~138, 2016.

\bibitem{qual4}
N.~Cruz, S.~Lepe, Y.~Leyva, F.~Pe\~na, and J.~Saavedra, ``{No stable
  dissipative phantom scenario in the framework of a complete cosmological
  dynamics},'' {\em Phys. Rev. D}, vol.~90, no.~8, p.~083524, 2014.

\bibitem{is2}
W.~Israel, ``{Nonstationary irreversible thermodynamics: A Causal relativistic
  theory},'' {\em Annals Phys.}, vol.~100, p.~310, 1976.

\bibitem{is2a}
D.~Pavon, ``{The Generalized second law and extended thermodynamics},'' {\em
  Class. Quant. Grav.}, vol.~7, p.~487, 1990.

\bibitem{is2b}
L.~Chimento and A.~S. Jakubi, ``{Cosmological solutions of the Einstein
  equations with a causal viscous fluid},'' {\em Class. Quant. Grav.}, vol.~10,
  p.~2047, 1993.

\bibitem{is3}
M.~Cruz, N.~Cruz, and S.~Lepe, ``{Accelerated and decelerated expansion in a
  causal dissipative cosmology},'' {\em Phys. Rev. D}, vol.~96, no.~12,
  p.~124020, 2017.

\bibitem{zeta}
I.~Brevik and O.~Gr\o{}n, ``{Universe Models with Negative Bulk Viscosity},''
  {\em Astrophys. Space Sci.}, vol.~347, pp.~399--404, 2013.

\bibitem{lepe1}
N.~Cruz, S.~Lepe, and F.~Pe\~na, ``{Crossing the phantom divide with
  dissipative normal matter in the Israel\textendash{}Stewart formalism},''
  {\em Phys. Lett. B}, vol.~767, p.~103, 2017.

\bibitem{lepe2}
M.~Cruz, N.~Cruz, and S.~Lepe, ``{Phantom solution in a non-linear
  Israel\textendash{}Stewart theory},'' {\em Phys. Lett. B}, vol.~769, p.~159,
  2017.

\bibitem{cataldo}
M.~Cataldo, N.~Cruz, and S.~Lepe, ``{Viscous dark energy and phantom
  evolution},'' {\em Phys. Lett. B}, vol.~619, pp.~5--10, 2005.

\bibitem{nonlinear}
M.~Cruz, N.~Cruz, E.~Gonz\'alez, and S.~Lepe, ``{Testing a non linear solution
  of the Israel-Stewart theory with supernovae},'' {\em arXiv:
  2109.08823[gr-qc]}, 2021.

\bibitem{libro}
J.~Wainwright and G.~F.~R. Ellis, {\em {Dynamical Systems in Cosmology}}.
\newblock Cambridge: Cambridge University Press, 1997.

\bibitem{qcoinci1}
W.~Zimdahl and D.~Pavon, ``{Interacting quintessence},'' {\em Phys. Lett. B},
  vol.~521, p.~133, 2001.

\bibitem{qcoinci2}
H.~Wei and S.~N. Zhang, ``{Observational $H(z)$ data and cosmological
  models},'' {\em Physics Lett. B}, vol.~644, p.~7, 2007.

\bibitem{planck}
N.~Aghanim {\em et~al.}, ``{Planck 2018 results. VI. Cosmological
  parameters},'' {\em Astron. Astrophys.}, vol.~641, p.~A6, 2020.

\bibitem{des}
T.~M.~C. Abbott {\em et~al.}, ``{Dark Energy Survey year 1 results:
  Cosmological constraints from galaxy clustering and weak lensing},'' {\em
  Phys. Rev. D}, vol.~98, no.~4, p.~043526, 2018.

\bibitem{kids}
H.~Hildebrandt {\em et~al.}, ``{KiDS-450: Cosmological parameter constraints
  from tomographic weak gravitational lensing},'' {\em Mon. Not. Roy. Astron.
  Soc.}, vol.~465, p.~1454, 2017.

\bibitem{valentino}
E.~D. Valentino, O.~Mena, S.~Pan, L.~Visinelli, W.~Yang, A.~Melchiorri, D.~F.
  Mota, A.~G. Riess, and J.~Silk, ``{In the realm of the Hubble
  tension{\textemdash}a review of solutions},'' {\em Class. Quant. Grav.},
  vol.~38, p.~153001, 2021.

\bibitem{dmde}
M.~Lucca and D.~C. Hooper, ``{Shedding light on dark matter-dark energy
  interactions},'' {\em Phys. Rev. D}, vol.~102, p.~123502, 2020.

\bibitem{halos}
P.~Carrilho, K.~Carrion, B.~Bose, A.~Pourtsidou, J.~C. Hidalgo, L.~Lombriser,
  and M.~Baldi, ``{On the road to~per\,cent accuracy VI: the non-linear power
  spectrum for interacting dark energy with baryonic feedback and massive
  neutrinos},'' {\em Mon. Not. Roy. Astron. Soc.}, vol.~512, p.~3691, 2022.

\bibitem{wang}
B.~Wang, E.~Abdalla, F.~Atrio-Barandela, and D.~Pavon, ``{Dark Matter and Dark
  Energy Interactions: Theoretical Challenges, Cosmological Implications and
  Observational Signatures},'' {\em Rept. Prog. Phys.}, vol.~79, p.~096901,
  2016.

\bibitem{lotka1}
J.~Perez, A.~F{\"u}zfa, T.~Carletti, L.~M{\'e}lot, and L.~Guedezounme, ``{The
  Jungle Universe: coupled cosmological models in a Lotka--Volterra
  framework},'' {\em General Relativity and Gravitation}, vol.~46, p.~1753,
  2014.

\bibitem{lotka2}
E.~Aydiner, ``{Chaotic universe model},'' {\em Sci. Rep.}, vol.~8, no.~1,
  p.~721, 2018.

\bibitem{lotka3}
M.~Cruz, S.~Lepe, and G.~Morales-Navarrete, ``{Qualitative description of the
  universe in the interacting fluids scheme},'' {\em Nucl. Phys.}, vol.~B,
  p.~114623, 2019.

\bibitem{magana}
J.~Magana, M.~H. Amante, M.~A. Garcia-Aspeitia, and V.~Motta, ``{The Cardassian
  expansion revisited: constraints from updated Hubble parameter measurements
  and type Ia supernova data},'' {\em Mon. Not. Roy. Astron. Soc.}, vol.~476,
  p.~1036, 2018.

\bibitem{mukherjee}
A.~Mukherjee, ``{Acceleration of the universe: a reconstruction of the
  effective equation of state},'' {\em Monthly Notices of the Royal
  Astronomical Society}, vol.~460, p.~273, 2016.

\bibitem{recent}
M.~Moresco, L.~Pozzetti, A.~Cimatti, R.~Jimenez, C.~Maraston, L.~Verde,
  D.~Thomas, A.~Citro, R.~Tojeiro, and D.~Wilkinson, ``{A 6\% measurement of
  the Hubble parameter at $z\sim0.45$: direct evidence of the epoch of cosmic
  re-acceleration},'' {\em JCAP}, vol.~05, p.~014, 2016.

\end{thebibliography}
\end{document}